\documentclass[twocolumn,preprintnumbers,amsmath,amssymb,floatfix]{revtex4}

\usepackage{graphicx}
\usepackage{dcolumn}
\usepackage{bm}
\usepackage{color}

\newcommand{\beq}{\begin{equation}}
\newcommand{\eeq}{\end{equation}}
\newcommand{\bey}{\begin{eqnarray}}
\newcommand{\eey}{\end{eqnarray}}

\begin{document}

\title{Possible formation of wormholes from dark matter in an isothermal galactic
     halo and void}

\author{Nayan Sarkar }
 \email{ nayan.mathju@gmail.com}
 \affiliation {Department of Mathematics, Jadavpur University, Kolkata-700032, India}

\author{Susmita Sarkar }
 \email{ susmita.mathju@gmail.com}
 \affiliation {Department of Mathematics, Jadavpur University, Kolkata-700032, India}

\author{Farook Rahaman}
\email{rahaman@associates.iucaa.in}
\affiliation {Department of Mathematics, Jadavpur University, Kolkata-700032, India}
\author{P.K.F. Kuhfittig}
\email{
kuhfitti@msoe.edu}
\affiliation {Department of Mathematics, Milwaukee School of Engineering, Milwaukee, Wisconsin
53202-3109, USA
}
\author{G S Khadekar}
\email{
gkhadekar@yahoo.com}
\affiliation {Department of Mathematics, R T M Nagpur University, Nagpur, India
}

\date{\today}

\begin{abstract}
It is well known that traversable wormholes are valid solutions of the Einstein field equations, but these structures can only be maintained by violating the null energy condition.  In this paper, we have obtained such wormhole solutions in an isothermal galactic halo, as well as in a void.  We have shown that the null energy condition is violated, with the help of a suitable redshift function obtained from flat galactic rotation curves.
\end{abstract}


\maketitle

\section{Introduction}

Wormholes are handles or tunnels in spacetime linking widely separated regions of our Universe or different Universes altogether.  Being such interesting phenomena, wormholes continue to be an active research topic in theoretical astrophysics. Einstein and Rosen had already considered a special type of solution from a physics point of view and became known as an Einstein-Rosen bridge connecting two identical spacetimes \cite{er35}.  Wheeler used $geons$  to construct the first diagram of a doubly-connected space on the Planck scale \cite{wh55}-\cite{wh62}.  He introduced the mysterious-sounding term $wormhole$ into the literature, while Hawking, in turn, transformed Wheeler wormholes into Euclidean wormholes \cite{ha88}.  Wheeler wormholes were not traversable, however, and could, in principle, develop some type of singularity \cite{rp67}.  Then in 1988,  Morris and Thorne proposed the first traversable wormhole \cite{mt88}; also, together with Yurtsever, they determined the energy conditions and even suggested the possibility of a time machine \cite{ty88}.

 According to the Standard Model of cosmology, the total mass-energy of the Universe consists of only 5\% ordinary matter, i.e., particles belonging to the standard model of particle physics.  The remaining 95\% of total mass-energy is contributed by the dark sector, which is thought to consist of dark matter (DM) and dark energy (DE). Roughly 27\% of the dark-sector component is the yet unseen dark matter, while dark energy, which drives the current cosmic acceleration, is responsible for the missing 68\%. Zwicky was the first astronomer to use the $ virial$ $ theorem $  to assert the existence of dark matter, which has been described as $dunkle$ $ Materie$ in galaxy clusters \cite{zw33}-\cite{zw37} since 1933.  The presence of DM in the galactic halo is deduced from its gravitational effect on the rotation curve of a spiral galaxy \cite{fr08}-\cite{tf06}.  That wormholes may occur naturally in the outer regions of a galactic halo supported by DM has already been considered in \cite{fr14} and \cite{pK14}.  These studies made use of the Navarro-Frenk-White (NFW) density profile \cite{NFW}.  It is well known that the NFW model produces constant tangential rotation curves in the outer regions.  By contrast, \cite{FR14} employs the Universal Rotation Curve (URC) dark-matter model to obtain analogous results for the central parts of the halo.

 Researchers have always been attracted to the study of wormholes in the framework of modified gravity and have subsequently succeeded in constructing wormhole solutions in different modified-gravity theories \cite{nf05}-\cite{mz16}. These include wormhole solutions in galactic halos, as exemplified by Sharif et al., who have obtained such solutions in the framework $f(T)$ teleparallel gravity \cite{ms14}, $f(G)$ gravity \cite{ms16}, and $f(G,T)$ gravity \cite{ms18}, the latter using the NFW density profile. That a dark-matter halo can support a traversable wormhole structure  in the outer region of spherical stellar systems is based on the Einasto density profile \cite{am16}. More recently, S. Islam et al. have proposed that the generalized Navarro-Frenk-White dark-matter density profile can support wormholes, unlike King's dark-matter profile, which fails to do so.  This model has been applied to the ultra diffuse galaxy (UDG) Dragonfly 44 in the Coma Cluster \cite{si18}, whose mass nearly equals that of the Milky Way galaxy and consists almost entirely of dark matter.

The main goal in this paper is to obtain wormhole solutions based on dark-matter models for an isothermal galactic halo and a void with the help of a suitable form of the redshift function obtained from flat rotation curves.

This paper is organized as follows: Sec. II presents the basic wormhole structure in terms of the so-called redshift and shape functions.  The Einstein field equations for the static, spherically symmetric line element are expressed in terms of these functions.  One form of the redshift function, obtained from the flat galactic rotation curves, is presented in Sec. III. In Sec. IV and Sec. V  we present the wormhole solutions corresponding to the isothermal density profile and the density profile of the void, respectively. Sec. VI discusses the junction to an external vacuum solution.  Finally, in Sec. VII we summarize the results.

\section{Wormhole structure and the Einstein field equations}

First we recall that a Morris-Thorne traversable  wormhole spacetime is represented by a static and spherically symmetric line element in Schwarzschild co-ordinates ($t$, $r$, $\theta$, $\phi$)\cite{mt88}:
\begin{equation}
ds^2 =- e^{2f(r)}dt^2+\left(1-\frac{b(r)}{r}\right)^{-1}dr^2+r^2(d\theta^2 + \text{sin}^2\theta \,d\phi^2)\label{metric},
\end{equation}\label{l}
where  $f=f(r)$ is the \emph{redshift function}, which must be everywhere finite to prevent an event horizon, and $b=b(r)$ is the \emph{shape function} since it helps determine the spatial shape of the wormhole.  The minimum radius $r=r_s$ in the metric coefficient $g_{rr}$ is the radius of the \emph{throat} of the wormhole, where $b(r_s) = r_s$ .  The shape function must satisfy the following conditions: $b(r)<r$ for $r>r_s$ and $b'(r_s)<1$, called the \emph{flare-out condition}. Finally, we require that $b(r)/r\rightarrow 0$ as $r\rightarrow\infty$.

Next, we recall the Einstein field equations
 \begin{equation}
 R_{\mu \nu}-{1\over 2}R~g_{\mu \nu}=8\pi T_{\mu \nu}, \label{eins}
\end{equation}
where $R_{\mu \nu}$, $g_{\mu \nu}$, $T_{\mu \nu}$, and $R$ are  the Ricci tensor, metric tensor, stress energy tensor, and Ricci scalar, respectively. Generally,  dark matter is represented by the  anisotropic energy momentum tensor

\begin{equation}
T_{\mu \nu} =  (\rho+p_t)U_\mu U_\nu-p_t g_{\mu \nu}+(p_r-p_t)\chi_\mu \chi_\nu, \label{2}
\end{equation}
where $U^\mu U_\mu = -\chi^\mu \chi_\mu = 1,~U^\mu \chi_\mu=0$; $\rho$, $p_r$, and $ p_t$ are  the energy density, radial pressure, and transverse pressure, respectively.

The Einstein field equations (\ref{eins}) now take on the following form for the metric (\ref{metric}) along with the energy momentum tensor (\ref{2}):

\begin{eqnarray}
8\pi\rho &=& \frac{b^\prime(r)}{r^2}, \label{den}\\
8\pi p_r &=& \left(1-\frac{b(r)}{r}\right)\left(\frac{1}{r^2}+\frac{2f^\prime(r)}{r}\right)-\frac{1}{r^2}, \label{pre}\\
8\pi p_t &=& \left(1-\frac{b(r)}{r}\right)f^{\prime\prime}(r)+\left(1-\frac{b(r)}{r}\right){f^\prime}^2(r)\nonumber
\\
&&+\frac{1}{2}\left(\frac{b(r)}{r^2}-\frac{b^\prime(r)}{r}\right)f^\prime(r)+\frac{1}{r}\left(1-\frac{b(r)}{r}\right)f^\prime(r)\nonumber
\\
&&+\frac{1}{2r}\left(\frac{b(r)}{r^2}-\frac{b^\prime(r)}{r}\right), \label{pt}
\end{eqnarray}

where $\prime$ stands for $\frac{d}{dr}$.

\begin{figure}[thbp]
\begin{center}
\begin{tabular}{rl}
\includegraphics[width=7.cm]{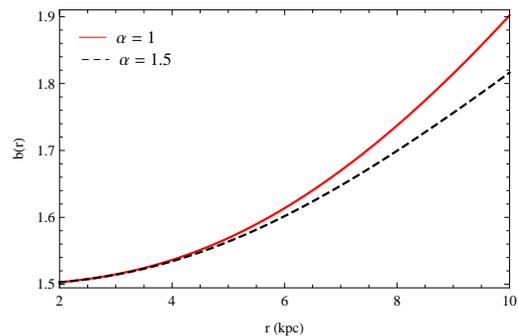}&
\\
\end{tabular}
\end{center}
\caption{ The shape function is plotted  with respect to the radial coordinate $r$ corresponding to $r_0 = 9.11$ kpc , $r_s = 1.5$ kpc , and $\rho_0 = 0.00008$.}\label{fig1}
\end{figure}

\section{The redshift function from flat rotation curves}

For the circular stable geodesic motion in the equatorial plane the tangential velocity can be obtained from the flat rotation curve \cite{ao16}; it is given by
\begin{equation}
(v^{\phi})^2 = rf^\prime\label{v}
\end{equation}
and fits the flat rotation curve for dark matter (DM). It is proposed by Rahaman et al.\cite{fr14, FR14} that the observed rotation curve profile in the dark-matter region is given by
\begin{equation}
v^{\phi}=\beta r e^{-kr}+\gamma[1-e^{-jr}]\label{vv},
\end{equation}
where $\gamma$, $\delta$, $k$, and $j$ are positive parameters.
Using Eq. (\ref{vv}) in Eq. (\ref{v}), yields the redshift function

\begin{eqnarray}
f&=&\gamma^2[\text{ ln}(r)+2E(1,jr)-E(1,2jr)]+D\nonumber
\\
&&-\frac{\beta^2e^{-2kr}}{2k }\left[r+\frac{1}{2k}\right]-2\beta\gamma e^{-kr}\left[\frac{1}{k}-\frac{1e^{-jr}}{(k+j) }\right]\nonumber,
\\
\label{f}
\end{eqnarray}

\begin{figure}[thbp]
\begin{center}
\begin{tabular}{rl}
\includegraphics[width=7.cm]{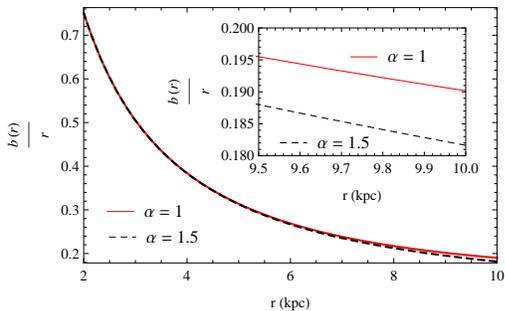}
\\
\end{tabular}
\end{center}
\caption{The function $\frac{b(r)}{r}$ is plotted with respect to the radial coordinate $r$ corresponding to $r_0 = 9.11$ kpc, $r_s = 1.5$ kpc, and $\rho_0 = 0.00008$.}\label{fig2}
\end{figure}
where $D$ is an integration constant and $E(n, x)$ is known as the exponential integral function, defined as
\begin{equation}
E(n,x)=E_n(x)=\int_0^1 e^{-\frac{x}{\eta}}\eta^{n-2}d\eta,
\end{equation}
where $ n = 0,1,2,\cdot\cdot\cdot$ and $x > 0.$

Next, we will use this redshift function to solve the Einstein field equations corresponding to two DM density profiles, the isothermal galactic halo and the void.

\begin{figure}[thbp]
\begin{center}
\begin{tabular}{rl}
\includegraphics[width=7.cm]{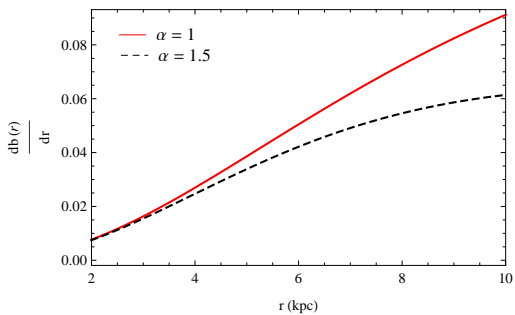}
\\
\end{tabular}
\end{center}
\caption{The function  $\frac{db(r)}{dr}$ is plotted with respect to the radial coordinate $r$ corresponding to $r_0 = 9.11$ kpc , $r_s = 1.5$ kpc, and $\rho_0 = 0.00008$.}\label{fig3}
\end{figure}


\section{A wormhole in the galactic halo}

 Let us consider the pseudo-isothermal density profile of the galactic halo
\begin{equation}
\rho(r)=\rho_0\left[1+\left({r\over r_0}\right)^2\right]^{-\alpha}\label{p},
\end{equation}
where $\rho_0$ and $r_0$ are the central density and core radius of the galactic halo, respectively, and $\alpha$ is a positive constant.
The given density profile for dark matter was first proposed by Kent\cite{ks86} in  1986 for $\alpha = 1$  and then extended in 2008 by M. Spano et al. to $\alpha = 1.5$\cite{ms08}. Accordingly, these two density profiles will be employed in our study.

\begin{figure}[thbp]
\begin{center}
\begin{tabular}{rl}
\includegraphics[width=7.cm]{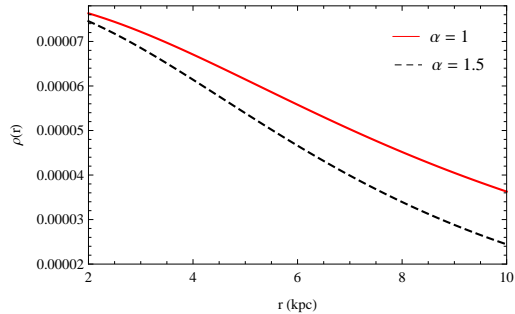}

\\
\end{tabular}
\end{center}
\caption{The energy density is plotted  with respect to the radial coordinate $r$ corresponding to $r_0 = 9.11$ kpc, $r_s = 1.5$ kpc, and $\rho_0 = 0.00008$.}\label{fig4}
\end{figure}

The exact behavior of this density function is shown in Fig. \ref{fig4}; it is monotonically decreasing, as expected.

After employing the above density profile (11) in  Eq. (\ref{den}) we  obtain the shape function

\begin{equation}
b(r)=\frac{8\pi \rho_0r^3}{3} F(r)+C,
\end{equation}
where $C$ is an integration constant and
\begin{equation}
F(r)=~~_2 F_1\left[\frac{3}{2},\alpha,\frac{5}{2},-\left(\frac{r}{r_0}\right)^2\right];
\end{equation} $_2F_1$ is the usual hypergeometric function defined as
\begin{eqnarray}
_2F_1 (a,~b;~c;~y) = \sum_{i=0}^\infty {(a)_i (b)_i \over (c)_i} {y^i \over i!}.
\end{eqnarray}
Here $(x)_n$ is the Pochhammer symbol, given by
\begin{eqnarray}
(x)_i = \left\{ \begin{array}{lcl}
1 & \mbox{for} & i=0 \\
x(x+1)....(x+i-1) & \mbox{for} & i>0 .
\end{array}\right.
\end{eqnarray}
Using the condition $b(r_s) = r_s$, we obtain the value of $C$:
\begin{equation}
C = r_s -\frac{8\pi \rho_0r_s^3}{3}F(r_s),
\end{equation}
where $r_s$ is the throat radius of the wormhole.
So the shape function becomes

\begin{eqnarray}
b(r)&=&\frac{8\pi \rho_0}{3}\bigg[r^3F(r)-r_s^3F(r_s)\bigg]+r_s\label{b}.
\end{eqnarray}

The check on the required conditions for the shape function was explored graphically.  Fig. \ref{fig1} shows the increasing behavior. The shape function is less than the radial coordinate $r$ for $r >r_s$, the throat radius $r_s = 1.5$ kpc, as can be seen from Fig. \ref{fig2}. Fig. 3 shows that the flare-out condition is satisfied by our shape function.

\begin{figure}[thbp]
\begin{center}
\begin{tabular}{rl}
\includegraphics[width=7.cm]{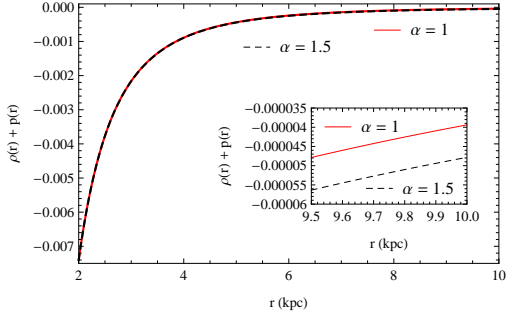}

\\
\end{tabular}
\end{center}
\caption{Variation of the null energy condition with respect to the radial coordinate $r$ corresponding to $r_0 = 9.11$ kpc, $j = k =1$, $ \beta= 0.001, \gamma= 0.1$, $r_s = 1.5$ kpc, and $\rho_0 = 0.00008$ .}\label{fig5}
\end{figure}


Using Eqs. (\ref{f}) and (\ref{b}) in Eq. (\ref{pre}), we get the expression for the radial pressure,

\begin{eqnarray}
p_r(r)&=&{1\over 24\pi k r^3}\left[k p_{r_1}+2\beta e^{-2(j+k)r}p_{r_2}p_{r_3}\right],\nonumber
\\
\end{eqnarray}
where
\begin{eqnarray}
p_{r_1}&=&3r_s+8\pi\rho_0 \left[r^3F(r)- r_s^3F(r_s)\right],\nonumber
\\
p_{r_2}&=&k\beta e^{2kr}\left(1+e^{2jr}-2e^{jr}\right)-2k\gamma re^{jr} e^{kr}\nonumber
\\
&&-re^{2jr}(kr\beta-\beta-4k\gamma),\nonumber
\\
p_{r_3}&=&3r-p_{r_1}.
\end{eqnarray}

Now recall that the violation of null energy condition is one of the essential conditions for the existence of a wormhole. It is showed in Fig. \ref{fig5} that the null energy condition is violated ($\rho + p_r < 0$) and hence the DM within the isothermal galactic halo satisfies the conditions for forming a wormhole.

\begin{figure}[thbp]
\begin{center}
\begin{tabular}{rl}
\includegraphics[width=7.cm]{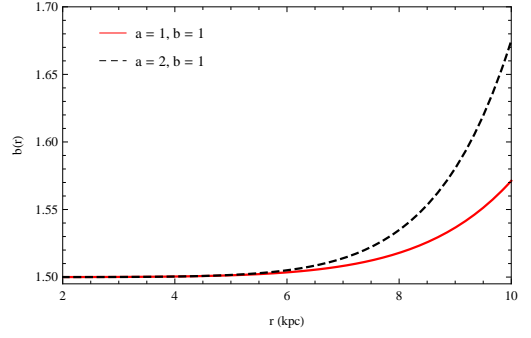}&
\\
\end{tabular}
\end{center}
\caption{The shape function $b(r)$ is plotted with respect to the radial coordinate $r$  corresponding to $r_0 = 3.5$ kpc, $r_s = 1.5$ kpc, and  $\rho_0 = 0.000001$.}\label{fig6}
\end{figure}

\section{A wormhole in the void}

The  density profile for the void is given by\cite{er14}
\begin{equation}
\rho(r)=\rho_0\left(\frac{r}{r_0}\right)^a \exp\left[{\left(\frac{r}{r_0}\right)^b-1}\right]\label{pp},
\end{equation}
where $\rho(r)$ is the density enclosed within the void-centric distance $r$, $\rho_0$ is the density enclosed within the void effective radius $r_0$, and
$a$ and $b$ are the best-fitting parameters. From the expression for the density it is obvious that the density within the void is always increasing with increasing radial coordinate $r$ (Fig. \ref{fig9}).
 In presenting the model, we will proceed with two pairs of values for $a$ and $b$.

\begin{figure}[thbp]
\begin{center}
\begin{tabular}{rl}
\includegraphics[width=7.cm]{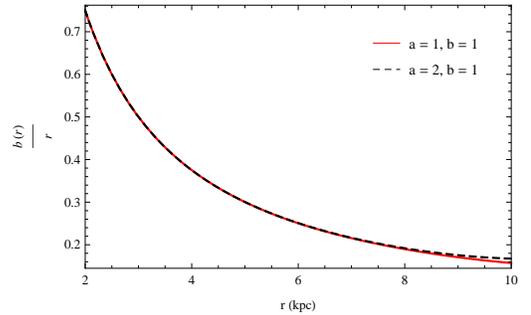}&
\\
\end{tabular}
\end{center}
\caption{The function   $\frac{b(r)}{r}$ is plotted with respect to the radial coordinate $r$   corresponding to  $r_0 = 3.5$ kpc and $r_s = 1.5$ kpc, with $\rho_0 = 0.000001$.}\label{fig7}
\end{figure}

 \subsection*{Case-I: For  $a = 1$, $b = 1$}
The void density profile (20)) for $a = b = 1$ becomes

\begin{equation}
\rho(r)=\rho_0\left(\frac{r}{r_0}\right) \exp\left[{\left(\frac{r}{r_0}\right)-1}\right]\label{pp1}.
\end{equation}

Using Eq. (\ref{pp1}) in  Eq. (\ref{den}), we obtain the shape function

\begin{equation}
b(r)=8\pi \rho_0\exp\left(\frac{r}{r_0}-1\right)\bigg[6rr_0^2-6r_0^3-3r^2r_0+r^3\bigg] +E,
\end{equation}
where $E$ is an integration constant obtained from $b(r_s) = r_s$ and is given by $  E = r_s-8\pi \rho_0\exp\left[\frac{r_s}{r_0}-1\right][6r_sr_0^2-6r_0^3-3r_s^2r_0+r_s^3]$, where  $r_s$ is the throat radius of the wormhole.  Therefore

\begin{eqnarray}
b(r)&=&8\pi \rho_0\bigg[\exp\left(\frac{r}{r_0}-1\right)\big[6rr_0-6r_0^3-3r^2r_0+r^3\big]\nonumber
\\
&&- \exp\left(\frac{r_s}{r_0}-1\right)\big[6r_sr_0^2-6r_0^3-3r_s^2r_0+r_s^3\big]\bigg]\nonumber
\\
& &+r_s\label{bv1}.
\end{eqnarray}

Using Eqs. (\ref{f}) and (\ref{bv1}) in Eq. (\ref{pre}), we obtain

\begin{eqnarray}
p_r&=&\frac{1}{8\pi kr^3}\bigg[2e^{-1-2(j+k)r}\beta p_{r_3}\bigg(e(r-r_0)-8\pi \rho_0 p_{r_4}\nonumber
\\
&&+e^{\frac{r_s}{r_0}}\big(r_s^3-3r_s^2r_0+6r_s r_0^2-6r_0^3\big)\bigg)\nonumber
\\
&&+k8\pi \rho_0\bigg(e^{\frac{r_s}{r_0}-1}\big(r_s^3-3r_s^2r_0+6r_s r_0^2-6r_0^3\big)\nonumber
\\
&&-e^{-1}p_{r_4}\bigg)-r_s\bigg],
\end{eqnarray}
where
\begin{eqnarray}
p_{r_3}&=&k\beta e^{2kr}\left[1+e^{2jr}-2e^{jr}\right]-re^{jr}\bigg[2k\gamma e^{kr}-e^{jr}\nonumber
\\
&&\times\{(kr-1)\beta-4k\beta\}\bigg],\nonumber
\\
p_{r_4}&=&e^{\frac{r}{r_0}}(r^3-3r^2r_0+6r r_0^2-6r_0^3\big).
\end{eqnarray}

\subsection*{Case-II: For  $a = 2$, $b = 1$}
Here we consider another pair of values of $a$ and $b$, namely $a = 2$ and $b = 1$. The density profile (20) becomes
\begin{equation}
\rho(r)=\rho_0\left(\frac{r}{r_0}\right)^2 \exp\left[{\left(\frac{r}{r_0}\right)-1}\right].\label{ppp}
\end{equation}

Applying a similar technique, we obtain the following expressions for the shape function and radial pressure, respectively:

 \begin{eqnarray}
b(r)&=&\frac{8\pi \rho_0}{r_0}\bigg[\exp\left(\frac{r}{r_0}-1\right)\big[24r_0^4-24rr_0^3+12r^2r_0^2\nonumber
\\
&&-4r^3r_0+r^4\big]- \exp\left(\frac{r_s}{r_0}-1\right)\big[24r_0^4-24r_sr_0^3\nonumber
\\
&&+12r_s^2r_0^2-4r_s^3r_0+r_s^4\big]\bigg]+r_s,\label{bv}
\end{eqnarray}

\begin{figure}[thbp]
\begin{center}
\begin{tabular}{rl}
\includegraphics[width=7.cm]{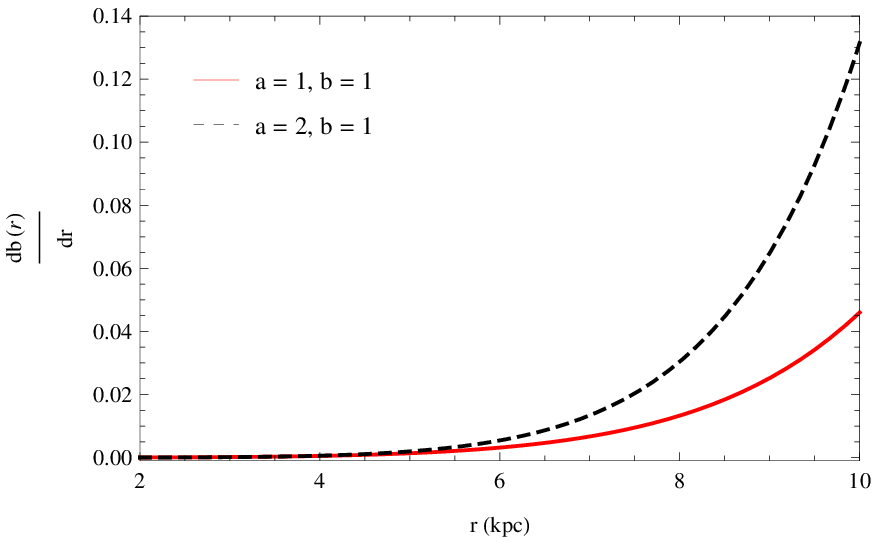}&
\\
\end{tabular}
\end{center}
\caption{The function   $\frac{db(r)}{dr}$ is plotted with respect to the radial coordinate $r$  corresponding to $r_0 = 3.5$ kpc and $r_s = 1.5$ kpc, with $\rho_0 = 0.000001$}\label{fig8}
\end{figure}

 \begin{eqnarray}
p_r&=&\frac{1}{8\pi k r_0 r^3}\bigg[2e^{-1-2(j+k)r}\beta p_{r_3}\bigg(er_0(r-r_0)-8\pi \rho_0 p_{r_5}\nonumber
\\
&&+e^{\frac{r_s}{r_0}}\big(24r_0^4-24r_s r_0^3+12r_s^2r_0^2-4r_s^3r_0+r_s^4\big)\bigg)\nonumber
\\
&&-k8\pi \rho_0\bigg(e^{-1}p_{r_5}-e^{\frac{r_s}{r_0}-1}\big(24r_0^4-24r_s r_0^3+12r_s^2r_0^2\nonumber
\\
&&-4r_s^3r_0+r_s^4\big)\bigg)-r_s\bigg],
\end{eqnarray}
where
\begin{eqnarray}
p_{r_5}&=&e^{\frac{r}{r_0}}\big(24r_0^4-24rr_0^3+12r^2r_0^2-4r^3r_0+r^4\big).
\end{eqnarray}
The check on the required conditions for the shape function was once again explored graphically in Figs. \ref{fig6}-\ref{fig8}.

\begin{figure}[thbp]
\begin{center}
\begin{tabular}{rl}
\includegraphics[width=7.cm]{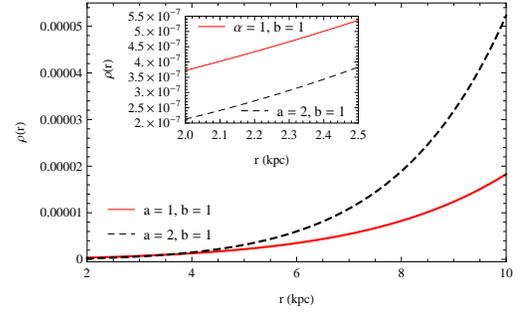}&
\\
\end{tabular}
\end{center}
\caption{Variation of the energy density  with respect to the radial coordinate $r$ corresponding to  $r_0 = 3.5$ kpc and $r_s = 1.5$ kpc, with $\rho_0 = 0.000001$.}\label{fig9}
\end{figure}


\begin{figure}[thbp]
\begin{center}
\begin{tabular}{rl}
\includegraphics[width=7.cm]{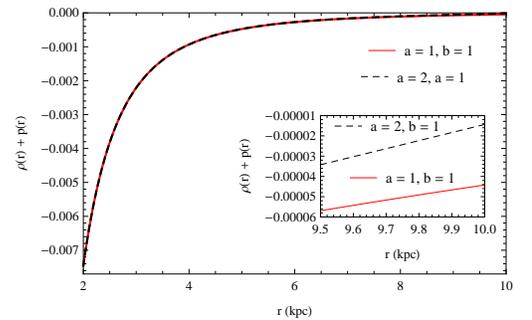}&
\\
\end{tabular}
\end{center}
\caption{Variation of the null energy condition  with respect to the radial coordinate $r$ corresponding to  $r_0 = 3.5$ kpc, $j = k =1$, $ \beta= 0.001, \gamma= 0.1$, $r_s = 1.5 $ kpc,  and $\rho_0 = 0.000001$.}\label{fig10}
\end{figure}

The null energy condition is  violated ($\rho + p_r < 0$) by our solution of the Einstein field equations corresponding to the density profile of the void, shown in Fig. \ref{fig10}, thereby satisfying the conditions required for the formation of a wormhole.

\section{Junction to an external vacuum solution}
As we have seen, $b(r)/r\rightarrow 0$ as $r\rightarrow\infty$, but $e^{2f(r)}$ does not approach unity, so that the wormhole spacetime described in line element (\ref{metric}) is not asymptotically flat.  The wormhole material must therefore be cut off at some $r_1>r_s$ and joined to an external vacuum solution.  Consider the Schwarzschild line element
\begin{multline}
ds^2 =- \left(1-\frac{2M}{r}\right)dt^2+\left(1-\frac{2M}{r}\right)^{-1}dr^2\\+r^2(d\theta^2 + \text{sin}^2\theta d\phi^2),
\end{multline}
where $M$ is the mass of the wormhole.  It follows directly that $\frac{1}{2}b(r_1)=M$.  Next, for the redshift function, Eq. (\ref{f}), we have to choose the arbitrary constant $D$ in such a way that $f(r_1)=\frac{1}{2}\text{ln}\,(1-2M/r_1)$.  Since $D$ is an additive constant, we simply let $r=r_1$ on the right-hand side of Eq. (\ref{f}), replace the left-hand side by $\frac{1}{2}\text{ln}\,(1-b(r_1)/r_1)$, and solve for $D$.  Then the line element becomes
\begin{multline}
ds^2 =- e^{2f(r)}dt^2+\left(1-\frac{b(r)}{r}\right)^{-1}dr^2\\+r^2(d\theta^2 + \text{sin}^2\theta \,d\phi^2),
   \quad\text{for}\quad r<r_1
\end{multline}
and
\begin{multline}
ds^2 =-\left(1-\frac{b(r_1)}{r}\right)dt^2+\left(1-\frac{b(r_1)}{r}\right)^{-1}dr^2\\+r^2(d\theta^2 + \text{sin}^2\theta \,d\phi^2),
   \quad\text{for}\quad r\ge r_1.
\end{multline}

\section{Results and discussion}
The possible existence and detection of wormholes in the galactic halo has already been discussed by means of the NFW density profile and the Universal Rotation Curve.  This paper uses two observationally motivated density profiles, the isothermal density profile
\begin{equation}
\rho(r)=\rho_0\left[1+\left({r\over r_0}\right)^2\right]^{-\alpha}\label{p}
\end{equation}
and the density profile for the void,
\begin{equation}
\rho(r)=\rho_0\left(\frac{r}{r_0}\right)^a \exp\left[{\left(\frac{r}{r_0}\right)^b-1}\right]\label{pp}.
\end{equation}
The density profiles are shown in Figs. \ref{fig4} and \ref{fig9}, respectively.  To obtain the wormhole solutions, we have determined a redshift function from the flat rotation curves in the dark-matter region proposed by Rahaman, et al.  The corresponding shape functions satisfy all the conditions for a traversable wormhole, as shown in Figs. \ref{fig1}-\ref{fig3} and Figs. \ref{fig6}-\ref{fig8}, respectively.  Figs. \ref{fig5} and \ref{fig10} show that the null energy condition is violated for the respective models.  So it seems entirely possible for naturally occurring wormholes to exist in both an isothermal galactic halo and in a void.

\section*{Acknowledgement}
Farook Rahaman would like to thank the authorities of the Inter-University Centre for Astronomy and
Astrophysics, Pune, India, for providing research facilities. Nayan Sarkar and Susmita Sarkar are  grateful to
CSIR (Grant No.: 09/096(0863)/2016-EMR-I.) and UGC (Grant No.: 1162/(sc)(CSIR-UGC
NET, DEC 2016)), Govt. of India, for financial support, respectively. G S Khadekar  is also thankful to  DST FIST PROGRAMME AND UGC SAP PROGRAMME for financial support.

\end{document}